# Temperature Dependence of Thermal Conductivity of Polycrystalline Graphene: Thermally Enhanced Kapitza Conductance

Young I. Jhon[†,1] and Myung S. Jhon[‡,1,2]

[1] Nano-convergence Core Technology for Human Interface (WCU), School of Advanced Materials Science and Engineering, Sungkyunkwan University, Suwon 440-746, Korea
[2] Department of Chemical Engineering, Carnegie Mellon University, Pittsburgh PA 15213, USA

**ABSTRACT:** Graphene has drawn wide attention due to its exceptional thermal conductivity but complete understanding of thermal characteristics of polycrystalline graphene is still elusive to date. For the first time, herein, we have systematically studied the effect of temperature on the thermal conductance behavior of polycrystalline graphene for a range of grain sizes and grain boundary types, by using non-equilibrium molecular dynamics simulations. It is noted that increasing the temperature remarkably enhances thermal boundary conductance (Kapitza conductance) across grain boundaries while it deteriorates thermal conductivities of defect-free grain regions, regardless of grain sizes and grain boundary types. This enhancement effect becomes more important for smaller grain sizes and higher temperatures, whose normal adverse effects on the heat transfer are thus ideally counterbalanced, making total thermal conductivity rather robust against the temperature increment for sufficiently small grain sizes. Upon heating from 300 to 500 K, thermal conductivity of graphene with maximally tilted grain boundaries decreases only by 9 % for a grain size of 50 nm in contrast to the decrease of 28 % for 250 nm, and further, it even increases below 30 nm. We presented quantitative mapping of its (grain, grain boundaries, and total) thermal conductivities in terms with grain sizes and temperatures, providing a guideline for graphene-based thermal engineering and suggesting novel defect-based thermal architectures as well.

## INTRODUCTION

With the aggressive downscaling of transistor technology toward nanometer scales and the concurrently increasing on-chip power consumption, thermal management in electronic circuits has become an integral part of the design for high-performance and reliable lifetimes of ultra large scale integrated (ULSI) systems.[1-3] The emergence of state-of-the-art recording technology, *i.e.*, heat assisted magnetic recording (HAMR) has also compounded this thermo-cooling issue as it demands operating temperature as high as 800 K.[4]

A possible approach for solving this problem is finding a material with an extremely high thermal conductivity.[5,6] This material can be integrated to CMOS devices and hard disk drive (HDD) employing HAMR technology, in order to prevent substrate temperature rises which may result in sub-threshold current leakage in the CMOS and critically damage the HDD. Besides pursuing efficient heat dissipations, acquiring the detailed information on system temperature distributions is also needed to establish a consolidated thermal management policy because complicated thermal distributions generally take place over the systems, creating interior hot spots.[7]

Graphene, a single layer of carbon atoms arranged in a honeycomb structure, is famous for many astonishing properties such as superb electrical conductivity and mechanical strength.[8-10] In particular, regarding thermal management solutions, it has attracted a great deal of attention due to its ultrafast thermal conductivity as well as excellent suitability for integration with CMOS devices compared to diamond and carbon nanotubes.[11-14] Recently, an innovative chemical vapor deposition technique was developed for the large-scale synthesis of graphene film,[15-18] opening a new avenue for the practical use of graphene in various fields. However, this method inherently generated the polycrystalline form of graphene due to the crystal imperfections of the substrate and the kinetics of the growth process,[19-22] and grain boundaries have become the most frequently observed defects in graphene.

A number of theoretical studies have been performed so far to characterize the thermal conductance of graphene in its various forms such as nanoribbion,[23,24] mono-layered,[25] few-layered,[26,27] and hydrogenated/hybrid structures.[28,29] However, the early studies have been dedicated to single crystalline graphene only, and the thermal properties of polycrystalline graphene have never been theoretically studied so far except for one latest pioneering study.[30] In this study, all investigations were performed assuming the systems were subject to room temperature, although the device temperature increments with a broad spectrum of spatial distribution would generally occur under actual operating conditions.

Here, to address this challenging issue, we have studied systematically the effect of temperature on thermal conductance behavior of polycrystalline graphene using non-equilibrium molecular dynamics (NEMD) simulations.[31,32] In this study, various zigzag and armchair-oriented tilted grain boundaries[33] were considered for different chiralities and misorientation angles, and the system temperature was changed from 300 to 700 K with intervals of 100 K. The thermal conductance of polycrystalline graphene was inspected separately for two characteristic regions, *i.e.*, grain boundaries and defect-free grain regions surrounded by grain boundaries.

The structures of the zigzag and armchair-oriented tilted grain boundaries employed in our study are depicted in Figure 1. The zigzag or armchair direction that is closest to perpendicular to the grain boundary is marked in magenta in Figures 1 (b) and (e). The chiral type of this direction determined whether the grain boundary is referred to as zigzag-oriented or armchair-oriented. The misorientation angle is defined as the angle made between this chiral direction and the normal direction of the grain boundary. For both zigzag and armchair-oriented grain boundaries, we examined three cases by decreasing the misorientation angle from the largest value and they were denoted by $ZZT_i$ and $ACT_i$ (i=1~3, in order from the largest to the smallest misorientation angle), respectively.

## COMPUTATIONAL METHODS

The thermal conductance was examined in the regime of the NEMD approach; when a heat flux is imposed through a polycrystalline system, temperature jumps emerge across grain boundaries and temperature gradients developed over grain regions, as illustrated in Figure 2. The boundary conductance (G), known as the Kapitza conductance,[34] is inversely proportional to the temperature jump ($\Delta T$) emerging across grain boundaries as given by

$$G = \frac{J}{\Delta T} \quad (1)$$

where $J$ is the heat flux, which is defined as the amount of energy per unit time transferred through each unit of cross sectional area.

In contrast, the thermal conductivity ($\lambda$) of grain regions is obtained by measuring the temperature gradient ($\triangledown T$) and using Fourier's law of

$$\lambda = \frac{J}{\triangledown T} \quad (2)$$

To obtain the temperature profile for $\triangledown T$, the systems were divided into slabs that were approximately 10 Å wide along the heat flow direction and their respective temperatures were calculated.

In the NEMD approach, imposing a heat flux is accomplished by periodically exchanging kinetic energies between coldest atoms located in the heat source region and hottest atoms located in the heat shrink region. After sufficient exchange has occurred, a steady-state heat flux is attained in the system, which can be calculated by

$$J = \frac{\Delta \varepsilon}{2 \Delta t A} \quad (3)$$

where $\Delta t$ is the time over which the simulation is performed, $\Delta \varepsilon$ is the total exchanged energy during the time period of $\Delta t$, and $A$ is the cross-sectional area perpendicular to the direction of the heat flow.

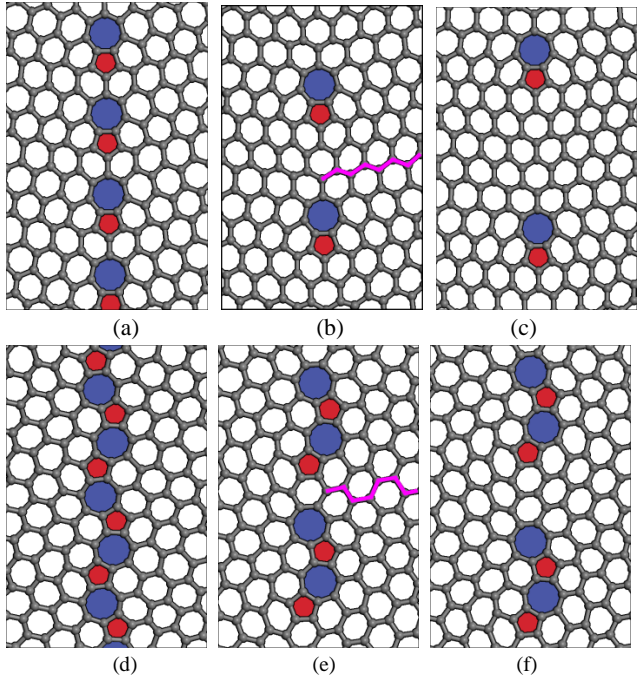

(a) (b) (c)
(d) (e) (f)

Figure 1. The structures of zigzag (top panel) and armchair (bottom panel) oriented tilted grain boundaries with (a,d) the first, (b,e) the second, and (c,f) the third largest misorientation angles. Specifically, the misorientation angles are 20.434°, 13.598°, and 11.138° for zigzag-oriented grain boundaries while they are 27.5°, 22.5°, and 18.3° for armchair-oriented ones.

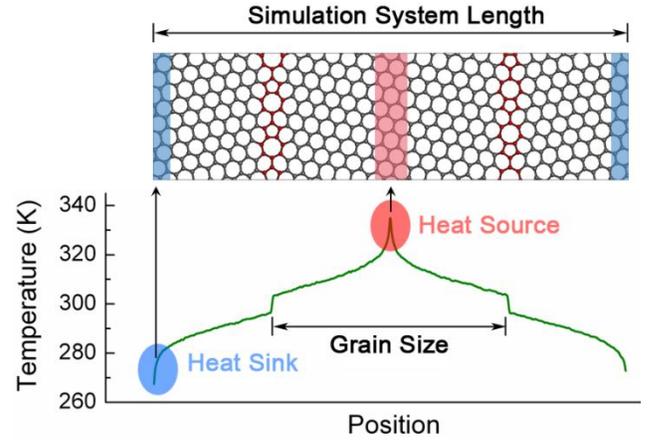

Figure 2. A schematic diagram of the simulation system for measuring the thermal conductivity of polycrystalline graphene in the regime of NEMD approach. The bonds marked in red denote grain boundaries which are directionally opposite to each other to fulfill the periodic boundary condition.

Three different dimensions of the systems were examined for each type of grain boundaries to see the (grain) size effect and they were approximately 100 nm, 250 nm, and 500 nm long along the heat flux direction with a common width of 50 nm, being composed of about $2\times10^4$, $5\times10^4$, and $1\times10^5$ carbon atoms, respectively. Periodic boundary conditions were applied to all systems by embedding two directionally opposite grain boundaries in the structures (Figure 2). The dimension of the simulation system and atomic coordinates were first optimized using a gradient-based minimization method with tolerance criteria of $10^{-8}$ eV/Å in force and/or $10^{-8}$ eV in energy. Based on the system size obtained above, the system was equilibrated with a time step of 0.5 fs for $1\times10^6$ steps at the desired temperature using a canonical NVT simulation. Then, with the imposition of heat flux, the canonical ensemble was performed consecutively for $2\times10^6$ steps to allow the system to reach the steady-state regime. Finally, a microcanonical NVE ensemble simulation was performed for $4\times10^6$ steps to calculate the thermal conductivity where the temperature profile and the heat flux were stored every 40 steps and averaged over the $1\times10^5$ samples. The simulations were performed using the LAMMPS (Large-scale Atomic/Molecular Massively Parallel Simulation) software package,[35] and the carbon atomic interaction was described by the optimized Tersoff potential proposed by Lindsay and Broido,[36] which has been shown to reproduce accurate acoustic phonon velocities in excellent agreement with experimental data. The acoustic phonon velocities contributed significantly to the thermal conductivity and in accordance with this fact, the Tersoff potential modification was shown to yield much improved thermal conductivity values for various carbon nanomaterials.[37,38]

## RESULTS AND DISCUSSION

We first calculated the thermal boundary conductance across zigzag-oriented grain boundaries at room temperature for various misorientation angles and grain sizes. We observed that the boundary conductance for these grain boundaries fell in the range of $2.0\times10^{10}$-$4.5\times10^{10}$ W/(m$^2$ K), in excellent agreement with the reported values of the previous study.[30] In addition, we have also calculated thermal conductivity in defect-free grain regions surrounded by grain boundaries. They were estimated to be 537.66±10.18, 987.07±16.51, and 1375.24±24.16 W/mK at room temperature for grain sizes of 50 nm, 100 nm, and 250 nm, respectively. These grain thermal conductivities were almost unaffected by the misorientation angle change of surrounding grain boundaries and were very similar to those (532, 898, and 1460 W/mK) of pristine graphene cells which have periodic lengths of 50 nm, 100 nm, and 250 nm along the heat flow direction, respectively.[30] These excellent coincidences definitely validate our computational approach for calculating the thermal conductance of polycrystalline graphene.

Next, the thermal boundary conductance of zigzag-oriented grain boundaries was plotted as a function of temperature to examine the temperature effect. Surprisingly, we found that the boundary conductance remarkably increased as the temperature increased, regardless of the misorientation angle and the grain size (Figures 3 (a) and S1). It is counterintuitive considering the normal adverse effect of temperature due to thermal perturbations. This trend was maintained at least up to 600-700K (few exceptional points were neglected in this statement considering the error tolerance). In contrast to such peculiar behaviors of thermal boundary conductance, the thermal conductivity of grain regions decreased as the temperature increased as is the case in pristine graphene (Figures 3 (b) and S2). We have also investigated the thermal conductance behavior of armchair-oriented grain boundaries, to our knowledge, which has never been inspected so far. They yielded smaller magnitudes of thermal boundary conductance compared to those of zigzag-oriented grain boundaries but showed very similar results generally (Figures 3 (c) and S1).

As the grain size decreased, both the thermal boundary conductance and the grain thermal conductivity decreased substantially for all grain boundaries types and temperatures. It is supposed to be due to a boundary effect shown in the system that is smaller than mean free path of phonons, which is exceptionally long for graphene as much as 775 nm.[39] As the misorientation angle decreased, the boundary conductance increased generally (Figure 4), while the thermal conductivity of grain regions remained almost constant. The extent of this increase was more striking between small misorientation angles (namely, for $T_2$ to $T_3$ rather than for $T_1$ to $T_2$) at low temperatures.

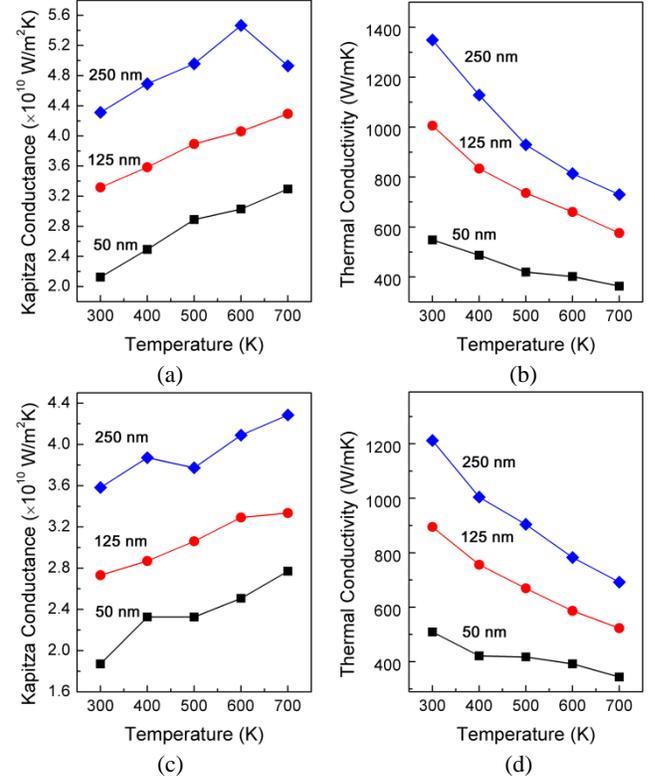

Figure 3. (a,c) Thermal boundary conductance of grain boundaries and (b,d) thermal conductivities of grain regions for various grain sizes and temperatures in polycrystalline graphene with $ZZT_1$ (top panel) and $ACT_1$ (bottom panel), respectively.

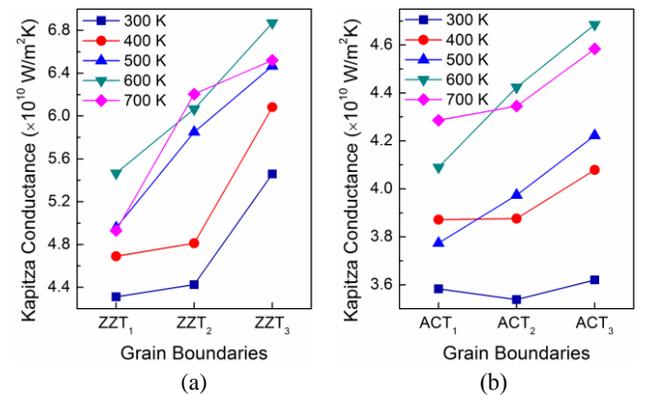

Figure 4. The thermal boundary conductance, known as the Kapitza conductance, plotted as a function of the misorientation angle at various temperatures for (a) zigzag-oriented and (b) armchair-oriented grain boundaries.

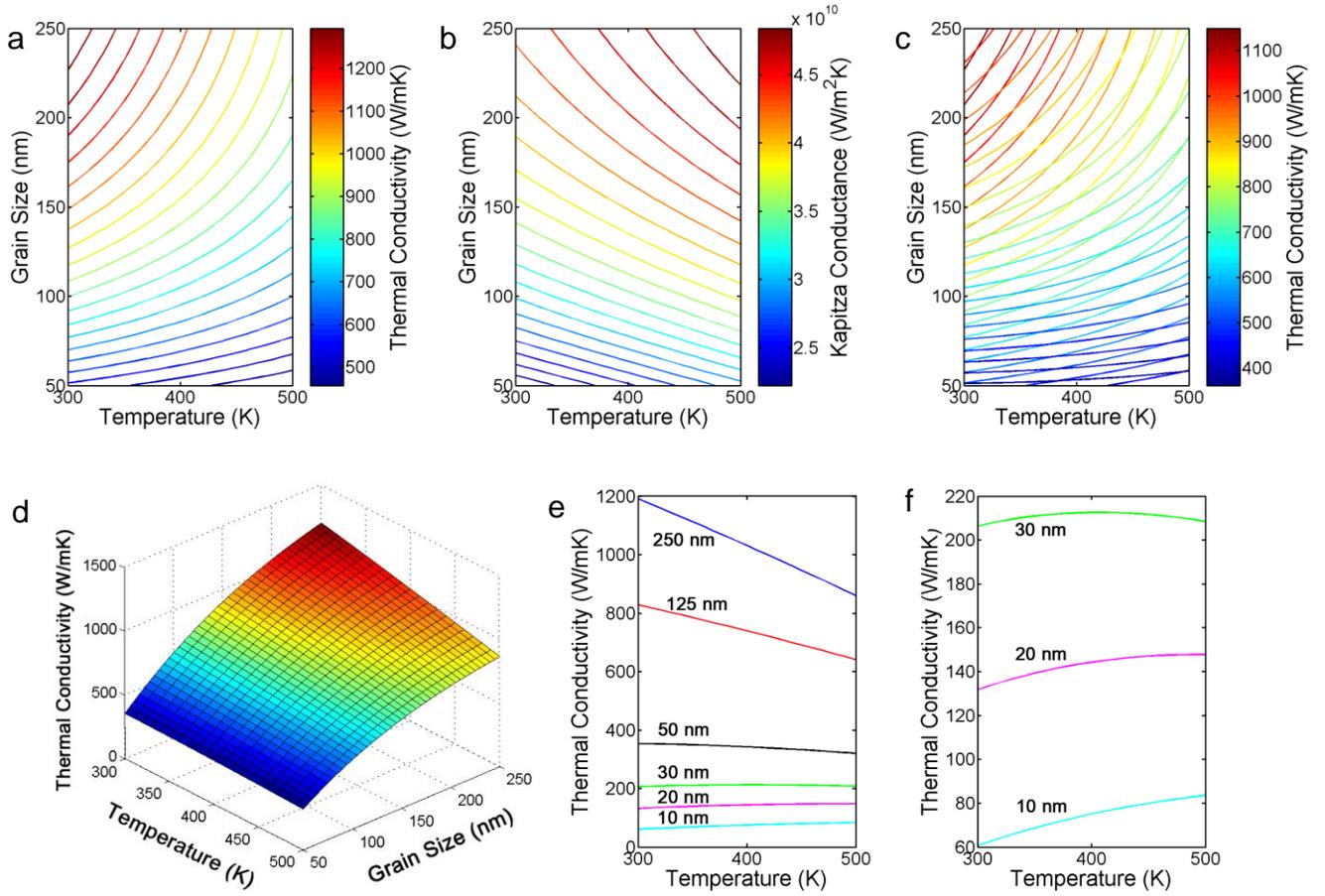

Figure 5. The contour lines of (a) the grain thermal conductivity ($\lambda_g$), (b) the thermal (grain) boundary conductance ($G$), and (c) the total thermal conductivity ($\lambda_t$) of polycrystalline graphene plotted as a function of the grain size and temperature. In Figure (c), the contour lines of $\lambda_g$ are also shown along with those of $\lambda_t$ for comparison, and we see that contour lines of $\lambda_t$ are more parallel to the temperature axis direction than those of $\lambda_g$, indicating its robustness against temperature variations. (d) The (total) thermal conductivity surface of polycrystalline graphene plotted as a function of the grain size and temperature. (e,f) The change of $\lambda_t$ is plotted as a function of temperature for various grain sizes.

The boundary conductance enhancement due to the temperature increase has been also observed in ultrathin diamond film,[40] several interfacial systems,[41] and multilayered structure[42] where the results were well explained by a theoretical model under the assumption of least interfacial scattering of phonons. Thus, we inferred that such a positive temperature dependence of thermal boundary conductance of polycrystalline graphene should be attributed to strong bond strengths and structural completeness (no dangling bonds) shown in the grain boundaries of graphene.

Under the heat flow imposed perpendicular to grain boundaries, the total thermal conductivity ($\lambda_t$) of polycrystalline graphene can be expressed using the boundary conductance ($G$) of grain boundaries, the thermal conductivity ($\lambda_g$) of grain regions, and the magnitude of the grain size ($l$), as written by

$$\lambda_t = [\lambda_g^{-1} + (lG)^{-1}]^{-1} \quad (4)$$

To investigate how the total thermal conductivity of polycrystalline graphene changes as the temperature and the grain size varies, we constructed quantitative mapping of these relations for $ZZT_1$ system as follows. We first derived the equations that can reproduce the simulation data of $\lambda_g(T, l)$ and $G(T, l)$ in the range of 300-500 K and 50-250 nm for $ZZT_1$ system, as written by

$$\lambda_g = a_1 + b_1 \times T, \ G = a_2 + b_2 \times T \quad (5)$$

Here, $a_1$, $b_1$, $a_2$, and $b_2$ are given in the following form of

$$[a_1, b_1, a_2, b_2] = c_1 \times \exp(-l/c_2) + c_3 \quad (6)$$

Their respective values of $c_1$, $c_2$, and $c_3$ are presented in Table S1. Using eqs 5-6, we first calculated the contour lines for $\lambda_g$ and $G$ as a function of the grain size ($l$) and temperature ($T$). For a specific grain size, the magnitude of $\lambda_g$ decreased as the temperature increased (Figure 5 (a)). For a specific temperature, the magnitude of $\lambda_g$ increased to presumably converge toward a certain (bulk) value as the grain size increased. This convergence occurred earlier for higher temperatures. On the other hand, the magnitude of $G$ increased as the temperature increased or the grain size increased (Figure 5 (b)).

Finally, using eqs 4-5, the total thermal conductivity ($\lambda_t$) was plotted as a function of the temperature ($T$) and the grain size ($l$) (Figures 5 (c) and (d)). We supposed that, as the grain size de-creases, the second term in the denominator of eq 4, which possesses positive temperature dependence should be dominate and thus, the resultant total thermal conductivity would be less deteriorated for the temperature increase because the second term would effectively counterbalance the first term that exhibits negative temperature dependence. We found that this conjecture is indeed correct, as seen in Figure 5 (c), where the contour lines of $\lambda_t$ are also shown along with those of $\lambda_g$ for comparison. Here, we see that the contour lines of $\lambda_t$ are more parallel to the direction of the temperature axis compared to those of $\lambda_g$, and its extent is more pronounced as the grain size decreases, indicating a robust heat transfer against temperature increase. Specifically, as the temperature increases from 300 K to 500 K, the thermal conductivity is reduced by 27.8% for a grain size of 250 nm while it decreases only by 9.1 % for a grain size of 50 nm (Figure 5 (e) and Table 1). Interestingly, the thermal conductivity is even predicted to increase for grain sizes below 30 nm as the temperature increases, provided that the extrapolation of the above equations are valid below a grain size of 50 nm (Figure 5 (f) and Table 1). In these peculiar cases, the thermal conductivity increases monotonically for grain sizes of 10 and 20 nm as the temperature increases, while it increases up to 400 K then decreases subsequently for a grain size of 30 nm.

Based on our above analyses, we also concluded that the presence of grain boundaries would never affect the thermal conductance behavior of graphene even at the higher temperatures (at least, up to 600~700 K) other than room temperature (this fact was already validated for room temperature[30]), provided that a grain size is larger than 1 μm, suggesting a promising potential of polycrystalline graphene for thermal cooling engineering.

**CONCLUSION**

In summary, the temperature effect on the thermal conductance behavior of polycrystalline graphene is systematically investigated using non-equilibrium molecular dynamics simulations. The thermal conductance is basically comprised of thermal boundary conductance across grain boundaries and thermal conductance in defect-free grain regions. Notably, as the temperature increases, the boundary conductance remarkably increases, presumably due to the great strengths and structural completeness of bonds comprising the grain boundaries, while the thermal conductivity of grain regions decreases due to thermal perturbations. The resultant total thermal conductivity is less sensitive (less deteriorated) to the temperature increase for smaller grain sizes and its extent is more pronounced as the grain size decreases. It is predicted that total thermal conductivity would even increase for grain sizes smaller than 30 nm.

The difference in the total thermal conductivity for different grain sizes is substantially reduced as the temperature increases. In addition, our work indicates that the presence of grain boundaries would not affect the thermal conductance behavior of graphene in the temperature range of 300K-700K provided that its grain size is larger than 1 μm, suggesting the promising potential of polycrystalline graphene for thermo-cooling applications. We believe that this study will greatly contribute to in-depth understanding of thermal conductance characteristics of polycrystalline as well as to the engineering for defect-based novel thermal architectures.

## Author information

**Corresponding Authors**
E-mail: yijhon@kaist.ac.kr (Y.I.J), mj3a@andrew.cmu.edu (M.S.J)

## Acknowledgement
This work was supported by the World Class University program of KOSEF (Grant No. R32-2008-000-10124-0).

## References
(1) Moore, G. E. Proc. of Inter. Elect. Device Meeting, 1975, 11-13.
(2) Naffziger, S.; Stackhouse B.; Grutkowski T.; Josephson, D.; Desai J.;Alon, E;Horowitz, M. IEEE J. of Solid-State Circuits 2006, 41 (1), 197-209.
(3) Borkar, S. IEEE Micro 1999, 19(4), 23–29.
(4) Mark, H. K.; Edward, C. G.; Terry, W. M.; William, A. C.; Robert, E. R.; Ganping, J.; Yiao, T. H.; Fatih E. Proc. of the IEEE 2008, 96(11), 1810-1835.
(5) Shen, S.;Henry, A.;Tong, J.;Zheng, R.;Chen, G. Nat. Nanotech. 2010, 5, 251-255.
(6) Balandin, A. A. Nat. Mater. 2011, 10, 569-581.
(7) Cohen, A. B.; Wang, P. J. of Heat Transfer, 2012, 134, 051017.
(8) Bolotin, K. I.; Sikes, K. J.; Jiang, Z.; Klima, M.; Fudenberg, G.; Hone, J.; Kim P.; Stormer, H. L. Solid State Commun. 2008, 146, 351-355.
(9) Lee, C; Wei, X; Kysar, J. W.; Hone. J. Science 2008, 321, 385-388.
(10) Liu, F.; Ming, P. M.; Li, J. Phys. Rev. B 2007, 76, 064120.
(11) Seol, J. H.; Jo, I.; Moore, A. L.; Lindsay, L.; Aitken, Z. H.; Pettes, M. T.; Li, X.; Yao, Z.; Huang, R.; Broido, D.; Mingo, N.; Ruoff, R. S.; Shi, L. Science 2010, 328, 213-216.
(12) Balandin, A. A.; Ghosh, S.; Bao, W.; Calizo, I.; Teweldebrhan, D.; Miao, F.; Lau, C. N. Nano Lett. 2008, 8, 902-907.
(13) Cai, W.; Moore, A. L.; Zhu, Y.; Li X.; Chen, S.; Shi, L.; Ruoff, R. S. Nano Lett. 2010, 10, 1645-1651.
(14) Ghosh, S.; Teweldebrhan, D.; Pokatilov, E. P.; Nika, D. L.; Balandin, A. A.; Bao, W.; Miao, F.; Lau C. N. Appl. Phys. Lett. 2008, 92, 151911.
(15) Yu, Q. K.; Lian, J.; Siriponglert, S.; Li, H.; Chen, Y. P.; Pei, S. S. Appl. Phys. Lett. 2008, 93, 113103.
(16) Kim, K. S.; Zhao, Y.; Jang, H.; Lee, S. Y.; Kim, J. M.; Kim, K. S.; Ahn, J. H.; Kim, P.; Choi, J. Y.; Hong, B. H. Nature, 2009, 457, 706-710.
(17) Li, X. S.; Cai, W. W.; An, J. H.; Kim, S.; Nah, J.; Yang, D. X.; Piner, R.; Velamakanni, A.; Jung, I.; Tutuc, E.; Banergee, S. K.; Colombo, L.; Ruoff, R. S. Science 2009, 324, 1312–1314.
(18) Levendorf, M. P.; Ruiz-Vargas C. S.; Garg, S.; Park, J. Nano Lett., 2009, 9, 4479–4483.

(19) Coraux, J.; N'Diaye A. T.; Engler, M.; Busse, C.; Wall, D.; Buckanie, N.; Meyer, F. J.; Heringdorf, Z.; Gastel, R. V.; Poelsema, B.; Michely, T. New J. Phys. 2009, 11, 023006.
(20) Loginova, E.; Nie, S.; Thurmer, K.; Bartelt, N. C.; McCarty, K. F. Phys. Rev. B 2009, 80, 085430.
(21) Miller, D. L.; Kubista, K. D.; Rutter, G. M.; Ruan, M.; de Heer, W. A.; First, P. N.; Stroscio, J. A. Science 2009, 324, 924-927.
(22) Park, H. J.; Meyer, J.; Roth, S.; Skakalova, V.; Carbon 2010 48, 1088-1094.
(23) Hu, J.; Ruan, X.; Chen, Y. P. Nano Lett. 2009, 9(7), 2730-2735.
(24) Tan, Z. W.; Wang, J.; Gan, C. K. Nano Lett. 2011, 11, 214-219.
(25) Zhang, H.; Lee, G.; Fonseca, A. F.; Borders, T. L.; Cho, K. J. of Nanomater. 2010, 537657.
(26) Lindsay, L.; Broido, D. A.; Mingo, N. Phys. Rev. B 2011, 83, 235428.
(27) Zhong, W. R.; Zhang, M. P.; Ai, B. Q; Zheng, D. Q. Appl. Phys. Lett. 2011, 98, 113107.
(28) Pei, Q. X.; Sha, Z. D.; Zhang, Y. W. Carbon 2011, 4752-4759.
(29) Varshney, V.; Patnaik, S. S.; Roy, A. K.; Froudakis, G.; Farmer, B. L. ACS Nano 2010, 4(2), 1153-1161.
(30) Bagri, A.; Kim, S. P.; Ruoff, R. S.; Shenoy, V. B. Nano Lett. 2011, 11, 3917-3921.
(31) MullerPlathe, F. J. Chem. Phys. 1997, 106 (14), 6082–6085.
(32) Muller-Plathe, F.; Reith, D. Comput. Theor. Polym. Sci. 1999, 9 (34), 203–209.
(33) Yazyev, O. L.; Louie, S. G. Phys. Rev. B 2010, 81, 195420.
(34) Schelling, P. K.; Phillpot, S. R.; Keblinski, P. J. Appl. Phys. 2004, 95 (11), 6082–6091.
(35) Plimpton, S. J. Comput. Phys. 1995, 117 (1), 1–19.
(36) Lindsay, L.; Broido, D. A. Phys. Rev. B 2010, 81(20), 205441.
(37) Lindsay, L.; Broido, D. A.; Mingo, N. Phys. Rev. B 2010, 82 (16), 161402.
(38) Haskins, J.; Kinaci, A.; Sevik, C.; Sevincli, H.; Cuniberti, G.; Cagin, T. ACS Nano 2011, 5(5), 3779–3787.
(39) Nika, D. L.; Askerov, A. S.; Balandin, A. A. Nano Lett. 2012, 12, 3238-3244.
(40) Angadi, M. A.; Watanabe, T.; Bodapati, A.; Xiao, X.; Auciello, O.; Carlisle, J. A.; Eastman, J. A.; Keblinski, P.; Schelling, P. K.; Phillpot, S. R. J. of Appl. Phys. 2006, 99, 114301.
(41) Samvedi, V.; Tomar V. Nanotech. 2009, 20, 365701.
(42) Cahill, D.G.; Goodson, K.; Majumdar, A. J. Heat Transfer 2002, 124, 223.

Table 1. The thermal conductivities of polycrystalline graphene for different temperatures and grain sizes. The unit for thermal conductivities is Watt per meter Kelvin (W/mK) and the values given in parentheses indicate the variation percentage to thermal conductivity values calculated at 300 K.

|       | 10 nm        | 20 nm         | 30 nm        | 50 nm         | 125 nm         | 250 nm          |
|-------|--------------|---------------|--------------|---------------|----------------|-----------------|
| 300 K | 60.6         | 131.7         | 206.4        | 345.8         | 794.6          | 1190.7          |
| 400 K | 75.1 (23.9 %)| 144.3 (9.6 %) | 212.6 (3.0 %)| 335.5 (-3.0 %)| 716.2 (-9.9 %) | 1030.7 (-13.4 %)|
| 500 K | 83.5 (37.8 %)| 147.8 (12.2 %)| 208.5 (1.0 %)| 314.5 (-9.1 %)| 626.5 (-21.2 %)| 859.9 (-27.8 %) |

Table of Contents

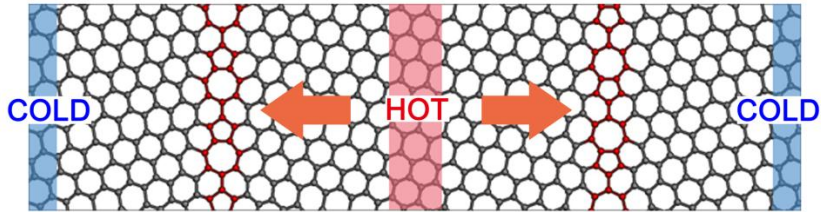
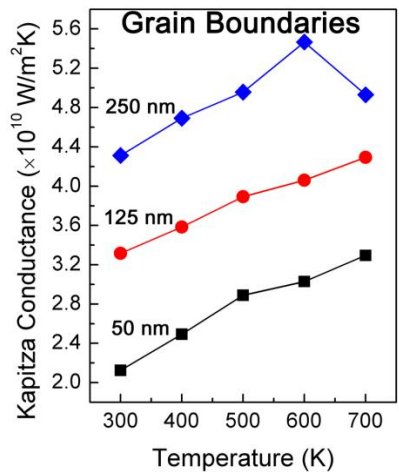 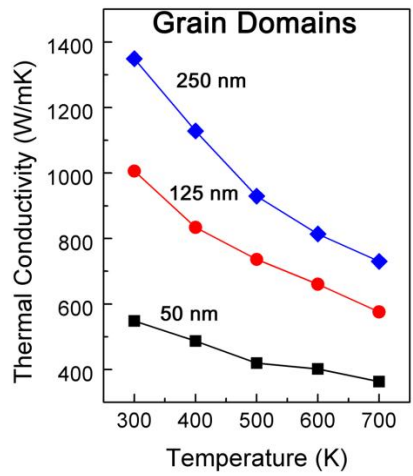

Supporting Online Material for

# Temperature Dependence of Thermal Conductivity of Polycrystalline Graphene: Thermally-Enhanced Kapitza Conductance

Young I. Jhon[1] and Myung S. Jhon[1,2]

[1] Nano-convergence Core Technology for Human Interface (WCU), School of Advanced Materials Science and Engineering, Sungkyunkwan University, Suwon 440-746, Korea

[2] Department of Chemical Engineering, Carnegie Mellon University, Pittsburgh PA 15213, USA

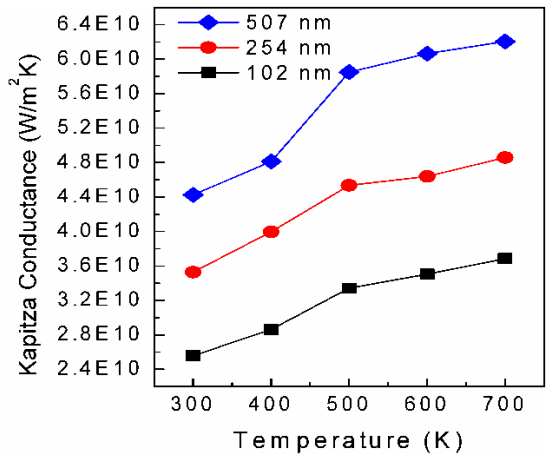
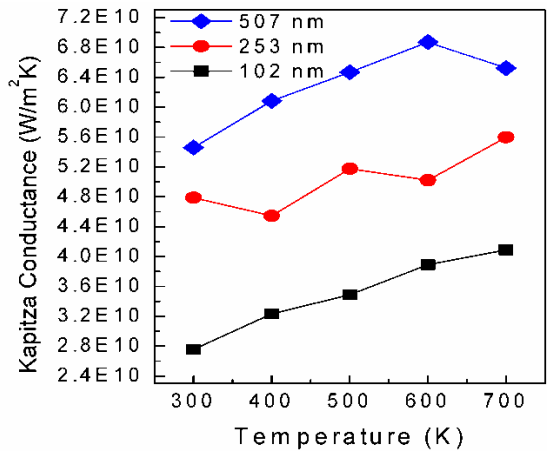

(a)

(b)

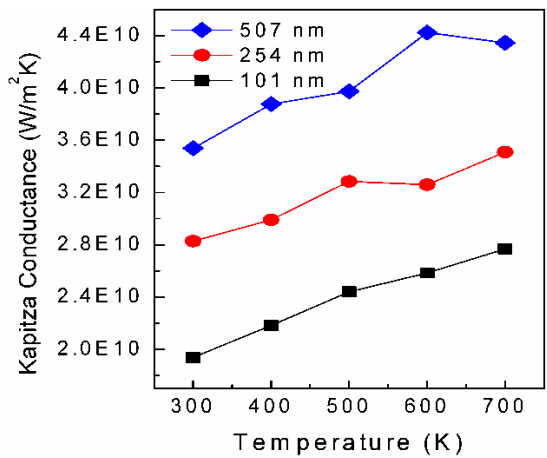
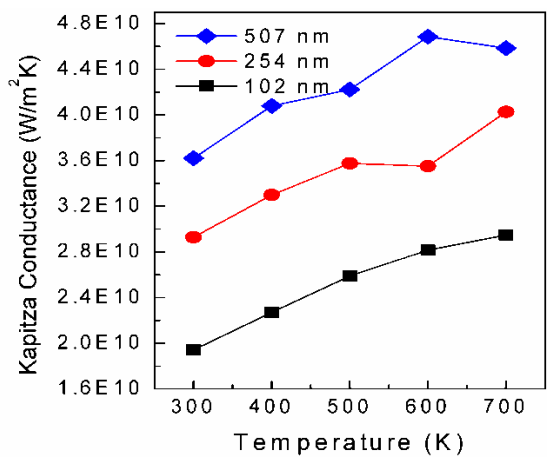

(c)

(d)

**Figure S1. (a-d)** The temperature-dependent Kapitza (thermal boundary) conductance of $ZZT_2$, $ZZT_3$, $ACT_2$, and $ACT_3$, respectively. They were studied for various grain sizes.

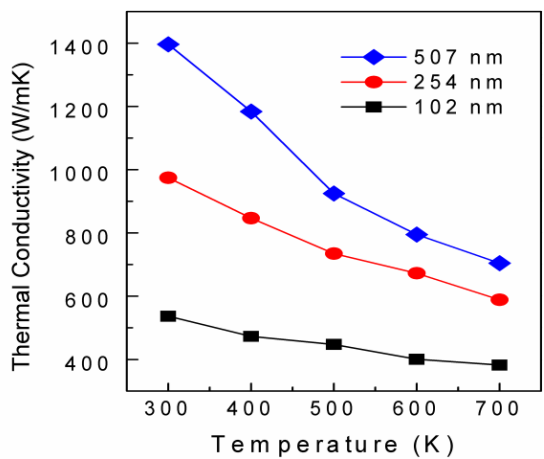
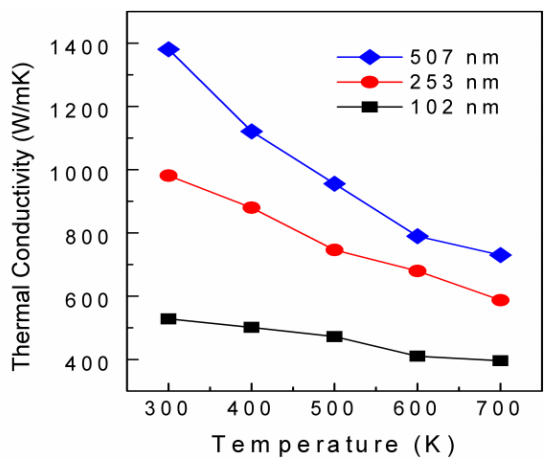
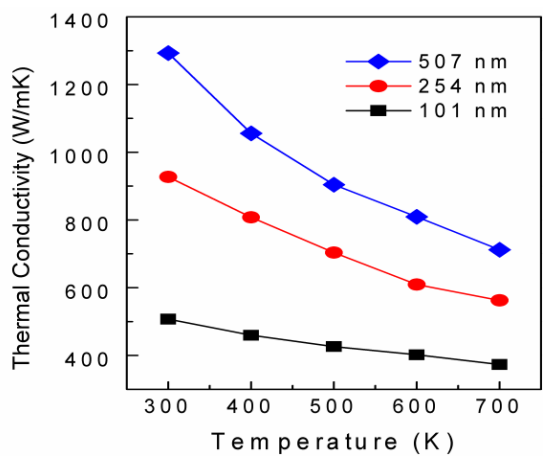
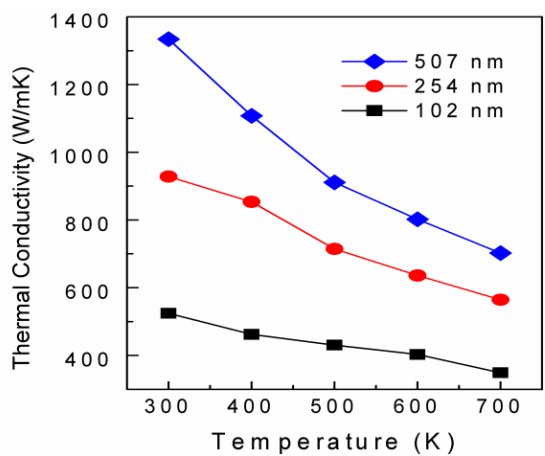

**Figure S2. (a-d)** The temperature-dependent thermal conductivities of grain regions of $ZZT_2$, $ZZT_3$, $ACT_2$, and $ACT_3$, respectively. They were studied for various grain sizes.

Table S1. The four sets of $c_1$, $c_2$, and $c_3$ values fitted to the parameters of $a_1$, $b_1$, $a_2$, and $b_2$, respectively, in the equations of $[a_1, b_1, a_2, b_2] = c_1 \times \exp(-l/c_2) + c_3$. The values for $a_1$, $b_1$, $a_2$, and $b_2$ had been obtained in advance through the fitting to the simulation data of Kapitza conductance ($G$) and grain thermal conductivity ($\lambda_g$) for the various grain sizes and temperatures, in the equations of $\lambda_g = a_1 + b_1 \times T$ and $G = a_2 + b_2 \times T$.

|       | $a_1$ | $b_1$ | $a_2$ | $b_2$ |
|-------|-------|-------|-------|-------|
| $c_1$ | -2376.04 (W/mK) | 3.105 (W/mK$^2$) | -4.59×10$^{10}$ (W/m$^2$K) | 1.34×10$^7$ (W/m$^2$K$^2$) |
| $c_2$ | 3181.96 (Å) | 4524.84 (Å) | 2067.00 (Å) | 1337.97 (Å) |
| $c_3$ | 2453.21 (W/mK) | -3.100 (W/mK$^2$) | 3.754×10$^{10}$ (W/m$^2$K) | 3.23×10$^7$ (W/m$^2$K$^2$) |